# Stabilization of Radiation Reaction with Vacuum Polarization

Keita Seto[1,*], Sen Zhang[2], James Koga[3],

Hideo Nagatomo[1], Mitsuo Nakai[1], Kunioki Mima[1,4]

[1]*Institute of Laser Engineering, Osaka University, 2-6 Yamada-oka, Suita, Osaka 565-0871, Japan*

[2]*Okayama Institute for Quantum Physics, Kyoyama 1-9-1, Kita-ku, Okayama 700-0015, Japan*

[3]*Quantum Beam Science Directorate, Japan Atomic Energy Agency, 8-1-7 Umemidai, Kizugawa, Kyoto 619-0215, Japan*

[4]*TheGraduate School for the Creation of New Photonics Industries, 1955-1 Kurematsu-cho, Nishi-ku, Hamamatsu, Shizuoka, 431-1202, Japan.*

*seto-k@ile.osaka-u.ac.jp

..................................................................

From the development of the electron theory by H. A. Lorentz in 1906, many authors have tried to reformulate this model. P. A. M. Dirac derived the relativistic-classical electron model in 1938, which is now called the Lorentz-Abraham-Dirac model. But this model has the big difficulty of the run-away solution. Recently, this equation has become important for ultra-intense laser-electron (plasma) interactions. For simulations in this research field, it is desirable to stabilize this model of the radiation reaction. In this paper, we will discuss this ability for radiation reaction with the inclusion of vacuum polarization.

..................................................................

1.  **Introduction**

With the rapid progress of ultra-short pulse laser technology, the maximum intensities of lasers have reached the order of $10^{22}$W/cm$^2$ [1, 2]. One laser facility which can achieve such ultra-high intensity is LFEX (Laser for fast ignition experiment) at the Institute of Laser Engineering (ILE), Osaka University [3] and another is the next generation laser project, the Extreme Light Infrastructure (ELI) project [4] in Europe. If an electron is in the strong fields caused by a laser which intensity is larger than $10^{18}$W/cm$^2$, the dynamics of the electron should be described by relativistic equations. The most important phenomenon in this regime is the effect of the ponderomotive force, where an electron is pushed in the propagation direction of the laser. When a charged particle is accelerated, it proceeds in a trajectory accompanied by bremsstrahlung. If the laser intensity is higher than $10^{22}$W/cm$^2$, which is referred to as ultra-relativistic laser intensities, strong bremsstrahlung might occur. Accompanying this, the "radiation reaction force (or damping force)" acts on the charged particle. Therefore, it is necessary to study radiation reaction effects in the ultra-relativistic laser-electron interaction regime. We can consider this effect as a "self-interaction". In laser-electron interactions at this intensity level, one of the important equations is the equation of motion including this reaction force.

However, in classical physics, these equations are based on H. A. Lorentz's work [5]. The purpose of his study was research into the electron's characteristics via classical physics. In particular, he was interested in radiation from a moving electron. The radiation field from a point electron is derived by the Liénard-Wiechert (retarded) potential in classical physics. Quantum mechanically, bremsstrahlung comes from the integrated momentum of all the photons emitted, meaning that an electron loses its kinetic energy via bremsstrahlung. However, classically an electron feels some force, which is related to bremsstrahlung, since there is an energy loss mechanism. Lorentz considered the electron model in which the charge is distributed on a sphere to investigate this force. His model was applied only in the non-relativistic regime, the case in which the electron has low velocity. One part of the electron interacts with another part by the Liénard-Wiechert electromagnetic field. Here, it is assumed, for analysis, all elements have the same velocity as the center of the electron. The Liénard-Wiechert field depends on the electron's motion and this field has directivity. If the electron is at rest, the Liénard-Wiechert field becomes the Coulomb repulsion. In this case, the electron has a spherically symmetrical field. Therefore, the electron doesn't feel a force apart from the Coulomb repulsion component of the Liénard-Wiechert field. However in the case with the moving electron, the radiation symmetry is broken by the directivity. This asymmetry leads to a self-interaction force from bremsstrahlung. This force is the "radiation reaction force".

After this work, P. A. M. Dirac updated Lorentz's theory with the relativistic covariant equation of motion, known as the Lorentz-Abraham-Dirac (LAD) equation in Minkowski spacetime [6]. Dirac suggested that the covariant form of the radiation reaction force should be described with, not only the retarded potential, but also with the advanced potential. When he suggested this theory, the theory of Quantum Electrodynamics (QED) had encountered a difficulty, a "mass divergence". It was resolved via the "renormalized theory" by S. Tomonaga [7], J. Schwinger [8], R. Feynman [9], F. Dyson [10] and so on. However, Dirac pursued physics, without the trick of mathematics. His hope was that the LAD equation would be a classical starting point to develop a quantum theory without the infinite self-energy of the electron [11]. In the end he was unable to achieve what he had hoped for. This method resulted in an equation, which is covariant. The starting point of discussion is this LAD equation as follows:

$$m_0 \frac{d}{d\tau} w^\mu = -e F_{\text{ex}}^{\mu\nu} w_\nu + f_{\text{LAD}}{}^\mu \qquad (1)$$

$$f_{\text{LAD}}{}^\mu = \frac{m_0 \tau_0}{c^2}\left(\frac{d^2 w^\mu}{d\tau^2} w^\nu - \frac{d^2 w^\nu}{d\tau^2} w^\mu\right) w_\nu \tag{2}$$

Where, $c\tau_0$ is the classical radius of the electron and $\tau_0 = e^2/6\pi\varepsilon_0 m_0 c^3$. $\tau$ is the proper time, $w$ is the 4-velocity defined as $w = \gamma(c, \mathbf{v})$. The Lorentz metric $g$ has the signature of $(+, -, -, -)$, $g_{\mu\nu} a^\mu a^\nu = a^\nu a_\nu = a^0 a^0 - a^1 a^1 - a^2 a^2 - a^3 a^3$. The 4-potential, the 4-current and the electromagnetic field tensor are defined by $A_{\text{ex}} = (\phi_{\text{ex}}/c, \mathbf{A}_{\text{ex}})$, $j = (c\rho, \mathbf{j})$ and $F_{\text{ex}}^{\mu\nu} = \partial^\mu A_{\text{ex}}{}^\nu - \partial^\nu A_{\text{ex}}{}^\mu$. Since these equations describe the effect of the radiation reaction, the LAD equations Eq.(1) and (2) become important in ultra-relativistic laser electron interactions. However, as was mentioned above, this equation has a run-away solution, which goes to infinity quickly (see [12, 13]). To avoid this run-away, there are many techniques from physical assumptions to mathematical treatments. Approximate methods have been suggested by Landau-Lifshitz [14] and Ford-O'Connell [15], F. Rohrlich [16], P. Caldirola [17] and I. V. Sokolov [18] suggested other methods, but we need to wait to decide which is correct in nature. Applications from fluid dynamics perspective were studied by K. K. Tam and D. Kiang [19] and V. I. Berezhiani *et al.* [20, 21].

However, there are vacuum polarizations in nonlinear QED. This knowledge leads us to a new electron model. In other words, it is possible to consider the problem of the radiation reaction with vacuum fluctuation via photon-photon scatterings. This vacuum polarization was investigated by Y. Monden and R. Kodama, under an ultra-intense laser in vacuum [22]. In this calculation, they used the formula of the "Heisenberg-Euler Lagrangian". The first order of this Lagrangian (SI units) is

$$L_{\text{Heisenberg-Euler}}^{\text{1st order}}\left(F_{\alpha\beta} F^{\alpha\beta}, F_{\alpha\beta}{}^* F^{\alpha\beta}\right) = -\frac{1}{4\mu_0} F_{\alpha\beta} F^{\alpha\beta} + \frac{\alpha^2 \hbar^3 \varepsilon_0^2}{360 m_0^4 c}\left[4\left(F_{\alpha\beta} F^{\alpha\beta}\right)^2 + 7\left(F_{\alpha\beta}{}^* F^{\alpha\beta}\right)^2\right]. \tag{3}$$

Here, $\alpha = e^2/4\pi\varepsilon_0 \hbar c$ is the fine structure constant. This formula was derived by W. Heisenberg and H. Euler [23], after that, J. Schwinger derived this by using gauge invariance [24]. This article follows this Lagrangian as quantum vacuum. At first, we will give a simple explanation about the Lorentz-Abraham-Dirac theory of an electron. And next, we will proceed to the vacuum polarization with the radiation from the electron. After that, we will discuss about the usability of this model of the radiation reaction with vacuum polarization, avoiding the run-away.

## 2. Dirac's Theory of Radiation Reaction

The Lorentz-Abraham-Dirac theory is the model of an electron in classical theory. Dirac considered the retarded and advanced field derived from the Maxwell equation. For a point electron ($x = x(\tau)$), these fields are as follows [6, 25]:

$$F_{\text{ret}}^{\mu\nu}\Big|_{x=x(\tau)} = \frac{3}{4}\frac{m_0\tau_0}{ec^2}\left(\frac{dw^\mu}{d\tau}w^\nu - w^\mu\frac{dw^\nu}{d\tau}\right)\int_{-\infty}^{\infty}d\delta\tau\frac{\delta(\delta\tau)}{|\delta\tau|} - \frac{m_0\tau_0}{ec^2}\left(\frac{d^2w^\mu}{d\tau^2}w^\nu - \frac{d^2w^\nu}{d\tau^2}w^\mu\right) \quad (4)$$

$$F_{\text{adv}}^{\mu\nu}\Big|_{x=x(\tau)} = \frac{3}{4}\frac{m_0\tau_0}{ec^2}\left(\frac{dw^\mu}{d\tau}w^\nu - w^\mu\frac{dw^\nu}{d\tau}\right)\int_{-\infty}^{\infty}d\delta\tau\frac{\delta(\delta\tau)}{|\delta\tau|} + \frac{m_0\tau_0}{ec^2}\left(\frac{d^2w^\mu}{d\tau^2}w^\nu - \frac{d^2w^\nu}{d\tau^2}w^\mu\right) \quad (5)$$

Both fields have a singular term, Dirac considered and defined the antisymmetrical field of which $F_{\text{LAD}}(x) = (F_{\text{ret}}(x) - F_{\text{adv}}(x))/2$. Specifically, for a point electron,

$$F_{\text{LAD}}^{\mu\nu}\Big|_{x=x(\tau)} = \frac{F_{\text{ret}}^{\mu\nu} - F_{\text{adv}}^{\mu\nu}}{2}\Big|_{x=x(\tau)} = -\frac{m_0\tau_0}{ec^2}\left(\frac{d^2w^\mu}{d\tau^2}w^\nu - \frac{d^2w^\nu}{d\tau^2}w^\mu\right), \quad (6)$$

and this field accelerates the electron itself,

$$f_{\text{LAD}}^\mu = -eF_{\text{reaction}}^{\mu\nu}w_\nu\Big|_{x=x(\tau)} = \frac{m_0\tau_0}{c^2}\left(\frac{d^2w^\mu}{d\tau^2}w^\nu - \frac{d^2w^\nu}{d\tau^2}w^\mu\right)w_\nu. \quad (7)$$

This then derives the LAD radiation reaction force in Eq.(1). J. Schwinger used the definition of Eq.(6), and obtained the equation of the energy loss of bremsstrahlung (the Larmor formula) [26]. This field, Eq.(6), is different from electromagnetic waves in vacuum. Electromagnetic waves in vacuum must satisfy the following relations:

$$F_{\text{in vacuum}\,\alpha\beta}F_{\text{in vacuum}}^{\alpha\beta} = 0 \quad (8)$$

$$F_{\text{in vacuum}\,\alpha\beta}{}^*F_{\text{in vacuum}}^{\alpha\beta} = 0 \quad (9)$$

Here, $^*F_{\text{in vacuum}}$ means the dual tensor of $F_{\text{in vacuum}}$. However, the LAD radiation reaction field satisfies,

$$F_{\text{LAD}\,\alpha\beta}F_{\text{LAD}}^{\alpha\beta} = 2\left(\frac{m_0\tau_0}{ec^2}\right)^2\left[c^2\frac{d^2w^\alpha}{d\tau^2}\frac{d^2w_\alpha}{d\tau^2} - \left(\frac{dw^\alpha}{d\tau}\frac{dw_\alpha}{d\tau}\right)^2\right], \quad (10)$$

$$F_{\text{LAD}\,\alpha\beta}{}^*F_{\text{LAD}}^{\alpha\beta} = 0, \quad (11)$$

where it does not always satisfy Eq.(8). When it doesn't satisfy Eq.(8) and Eq.(9), it must be regarded that the LAD field is not an electromagnetic wave propagating in vacuum. Therefore, the LAD field is nonlinear.

3. **Radiation Reaction with Quantum Vacuum fluctuation**

In this section, we will proceed to another scheme for radiation reaction by using the process of the vacuum polarization (friction). The Heisenberg-Euler Lagrangian is for dynamics in the low frequency approximation ($\hbar\omega \ll mc^2$). Therefore, we need to investigate the full Lagrangian of quantum vacuum, $L_{\text{Quantum Vacuum}}(F_{\alpha\beta}F^{\alpha\beta}, F_{\alpha\beta}{}^*F^{\alpha\beta})$ for general fields $F$, which converges to the Heisenberg-Euler Lagrangian for a constant field. However, its concrete form hasn't been derived yet. For generalization, a correct formalism in high-order terms is required. Therefore, it is considered that the Lagrangian $L_{\text{quantum vacuum}}$ includes $L_{\text{Heisenberg-Euler}}^{\text{1st order}}$, as the lowest order component. In accordance with this idea, since $L_{\text{Quantum Vacuum}}(F_{\alpha\beta}F^{\alpha\beta}, F_{\alpha\beta}{}^*F^{\alpha\beta}) = L_{\text{Heisenberg-Euler}}^{\text{1st order}}(F_{\alpha\beta}F^{\alpha\beta}, F_{\alpha\beta}{}^*F^{\alpha\beta}) + \cdots$ for the general field $F$, we will use the lowest-order Heisenberg-Euler Lagrangian as the quantum vacuum. From Eq.(10), the reaction field at the electron's point is scattered slightly by the quantum vacuum. For this formulation, we need to consider photon-photon scattering in nonlinear QED. From the Lagrangian Eq.(3), the Maxwell's equations are derived,

$$\partial_\mu \left[ F^{\mu\nu} + \frac{1}{\varepsilon_0 c} M^{\mu\nu} \right] = 0 \tag{12}$$

$$M^{\mu\nu} = -\frac{\alpha^2 \hbar^3 \varepsilon_0^2}{45 m_0^4 c^2} \left[ 4\left(F_{\alpha\beta}F^{\alpha\beta}\right)F^{\mu\nu} + 7\left(F_{\alpha\beta}{}^*F^{\alpha\beta}\right)^*F^{\mu\nu} \right] \tag{13}$$

One of the solutions of Eq.(12) is

$$F^{\mu\nu}(x) = -\frac{1}{\varepsilon_0 c} M^{\mu\nu}(x) + \tilde{C}^{\mu\nu}(x), \tag{14}$$

with $\tilde{C}^{\mu\nu}(x)$ satisfying the relation $\partial_\mu \tilde{C}^{\mu\nu}(x)=0$. $M^{\mu\nu}$ in Eq.(14), means the polarization of the vacuum fluctuation. If this fluctuation doesn't exist, it becomes in the classical limit of $\hbar \to 0$, the field $F^{\mu\nu}(x)$ needs to connect to the previous LAD theory. Then, $F^{\mu\nu}(x) = \tilde{C}^{\mu\nu}(x) = F_{\text{LAD}}{}^{\mu\nu}(x)$,

$$F^{\mu\nu} + \frac{1}{\varepsilon_0 c} M^{\mu\nu} = F_{\text{LAD}}{}^{\mu\nu}. \tag{15}$$

The relation $\tilde{C}^{\mu\nu}(x) = F_{\text{LAD}}{}^{\mu\nu}(x)$ is supported by,

$$\partial_\mu \frac{F_{\text{ret}}{}^{\mu\nu}(x) - F_{\text{adv}}{}^{\mu\nu}(x)}{2} = 0, \tag{16}$$

on any $x$. Our model requires that Eq.(15) holds at electron position $x(\tau)$, so from now on we will drop the subscript " $x = x(\tau)$ " in our calculations. From Eq.(15) with Eq.(14),

$$F^{\mu\nu} = F_{\text{LAD}}{}^{\mu\nu} + \eta \left[ \left(F_{\alpha\beta}F^{\alpha\beta}\right)F^{\mu\nu} + \frac{7}{4}\left(F_{\alpha\beta}{}^*F^{\alpha\beta}\right)^*F^{\mu\nu} \right]. \tag{17}$$

( $\eta = 4\alpha^2 \hbar^3 \varepsilon_0 / 45 m_0^4 c^3 = 2.6 \times 10^{-24}$ ). By perturbation of Eq.(17) near $\eta = 0$, it becomes $F = \sum_{n=0}^{\infty} \eta^n q_n F_{\text{LAD}}$, here $q_{n=0,1,2,\cdots} \in \mathbb{R}$, generally,

$$F^{\mu\nu} = F_{\text{LAD}}{}^{\mu\nu} + \delta F_{\text{LAD}}{}^{\mu\nu} = (1+\delta) F_{\text{LAD}}{}^{\mu\nu}, \tag{18}$$

because $F_{\text{LAD}\,\alpha\beta}{}^{*}F_{\text{LAD}}{}^{\alpha\beta}=0$ and $M^{\mu\nu}/\varepsilon_0 c \ll F_{\text{LAD}}{}^{\mu\nu}$. Here, the small value $\delta \in \mathbb{R}$ will be defined by $\delta \times F^{\mu\nu} = -M^{\mu\nu}/\varepsilon_0 c$. This leads to $F_{\alpha\beta}{}^{*}F^{\alpha\beta}=0$, Eq.(17) becomes,

$$F^{\mu\nu} = \frac{1}{1-\eta F_{\alpha\beta}F^{\alpha\beta}} F_{\text{LAD}}{}^{\mu\nu}. \tag{19}$$

From Eq.(18) and Eq.(19), using the scale in which $\delta \ll 1$,

$$\begin{aligned}
F^{\mu\nu} &= \frac{F_{\text{LAD}}{}^{\mu\nu}}{1-\eta(1+\delta)^2\left(F_{\text{LAD}\,\alpha\beta}F_{\text{LAD}}{}^{\alpha\beta}\right)} \\
&= \frac{F_{\text{LAD}}{}^{\mu\nu}}{1-\eta\left(F_{\text{LAD}\,\alpha\beta}F_{\text{LAD}}{}^{\alpha\beta}\right)-\eta\left(\delta^2+2\delta\right)\left(F_{\text{LAD}\,\alpha\beta}F_{\text{LAD}}{}^{\alpha\beta}\right)} \\
&= \frac{F_{\text{LAD}}{}^{\mu\nu}}{1-\eta\left(F_{\text{LAD}\,\alpha\beta}F_{\text{LAD}}{}^{\alpha\beta}\right)} \frac{1}{1-\dfrac{\eta\delta(\delta+2)\left(F_{\text{LAD}\,\alpha\beta}F_{\text{LAD}}{}^{\alpha\beta}\right)}{1-\eta\left(F_{\text{LAD}\,\alpha\beta}F_{\text{LAD}}{}^{\alpha\beta}\right)}} \\
&= \frac{F_{\text{LAD}}{}^{\mu\nu}}{1-\eta\left(F_{\text{LAD}\,\alpha\beta}F_{\text{LAD}}{}^{\alpha\beta}\right)} \sum_{n=0}^{\infty}\left[\frac{\eta\delta(\delta+2)\left(F_{\text{LAD}\,\alpha\beta}F_{\text{LAD}}{}^{\alpha\beta}\right)}{1-\eta\left(F_{\text{LAD}\,\alpha\beta}F_{\text{LAD}}{}^{\alpha\beta}\right)}\right]^n.
\end{aligned} \tag{20}$$

Therefore, from Eq.(18, 20) and $|\delta| \ll 1$,

$$\delta = \frac{\eta\left(F_{\text{LAD}\,\alpha\beta}F_{\text{LAD}}{}^{\alpha\beta}\right)}{1-\eta\left(F_{\text{LAD}\,\alpha\beta}F_{\text{LAD}}{}^{\alpha\beta}\right)}\left\{1+\frac{\delta(\delta+2)}{1-\eta\left(F_{\text{LAD}\,\alpha\beta}F_{\text{LAD}}{}^{\alpha\beta}\right)}\sum_{n=1}^{\infty}\left[\frac{\eta\delta(\delta+2)\left(F_{\text{LAD}\,\alpha\beta}F_{\text{LAD}}{}^{\alpha\beta}\right)}{1-\eta\left(F_{\text{LAD}\,\alpha\beta}F_{\text{LAD}}{}^{\alpha\beta}\right)}\right]^{n-1}\right\}$$

$$\sim \eta\left(F_{\text{LAD}\,\alpha\beta}F_{\text{LAD}}{}^{\alpha\beta}\right). \tag{21}$$

Then the field $F$ in Eq.(20) becomes,

$$F^{\mu\nu} = \frac{1+\mathrm{O}\left((\eta F_{\text{LAD}\,\alpha\beta}F_{\text{LAD}}{}^{\alpha\beta})^2\right)}{1-\eta\left(F_{\text{LAD}\,\alpha\beta}F_{\text{LAD}}{}^{\alpha\beta}\right)} F_{\text{LAD}}{}^{\mu\nu}. \tag{22}$$

Then, the equation of motion without an external field is

$$m_0\left[1-\eta\left(F_{\text{LAD}\,\alpha\beta}F_{\text{LAD}}{}^{\alpha\beta}\right)\right]\frac{d}{d\tau}w^\mu = -e\left[1+\mathrm{O}\left((\eta F_{\text{LAD}\,\alpha\beta}F_{\text{LAD}}{}^{\alpha\beta})^2\right)\right]F_{\text{LAD}}{}^{\mu\nu}w_\nu. \tag{23}$$

When the external field exists,

$$m_0\left[1-\eta\left(F_{\text{LAD}\,\alpha\beta}F_{\text{LAD}}{}^{\alpha\beta}\right)\right]\frac{d}{d\tau}w^\mu = -eF_{\text{ex}}^{\mu\nu}w_\nu$$

$$-e\left[1+\mathrm{O}\left((\eta F_{\text{LAD}\,\alpha\beta}F_{\text{LAD}}{}^{\alpha\beta})^2\right)\right]F_{\text{LAD}}{}^{\mu\nu}w_\nu. \tag{24}$$

By neglecting $\mathrm{O}\left((\eta F_{\text{LAD}\,\alpha\beta}F_{\text{LAD}}{}^{\alpha\beta})^2\right)$, we can obtain the corrected equation,

$$m_0\frac{d}{d\tau}w^\mu = -\frac{e}{1-\eta\left(F_{\text{LAD}\,\alpha\beta}F_{\text{LAD}}{}^{\alpha\beta}\right)}\left(F_{\text{ex}}^{\mu\nu}+F_{\text{LAD}}{}^{\mu\nu}\right)w_\nu. \tag{25}$$

Here, $F_{LAD\,\alpha\beta} F_{LAD}{}^{\alpha\beta} \leq 0$, this new field doesn't have any singular point. This equation is based on the Heisenberg-Euler Lagrangian, therefore, it requires us to satisfy that the LAD field must be under the Schwinger limit. Since we consider only the lowest order Heisenberg-Euler Lagrangian, the neglected terms in Eq.(22) should be considered when investigating using high order terms of the Heisenberg-Euler Lagrangian.

4. **Run-away Avoidance**

The new equation, Eq.(25) is required to avoid the run-away. The run-away means the self-acceleration by the infinite emission of light. Of course, this is a serious problem, which we should suppress. Here, we will discuss this problem. When the dynamics becomes run-away in the case of a non-external field,

$$\frac{d}{d\tau}w^\mu = -\frac{e}{m_0}\frac{F_{LAD}{}^{\mu\nu}w_\nu}{1-\eta\left(F_{LAD\,\alpha\beta} F_{LAD}{}^{\alpha\beta}\right)} \to \infty. \qquad (26)$$

From this equation of motion (26),

$$-m_0\tau_0 \frac{dw^\mu}{d\tau}\frac{dw_\mu}{d\tau} = \frac{\tau_0(ec)^2}{2\eta m_0} \frac{\left(-\eta F_{LAD\,\alpha\beta} F_{LAD}{}^{\alpha\beta}\right)}{\left[1+\left(-\eta F_{LAD\,\alpha\beta} F_{LAD}{}^{\alpha\beta}\right)\right]^2} \qquad (27)$$

where the relation that $w^\nu F_{LAD\,\mu\nu} F_{LAD}{}^{\mu\nu} w_\nu = e^2 c^2/2 \times F_{LAD\,\alpha\beta} F_{LAD}{}^{\alpha\beta}$ was applied. The right hand side has the factor of $x/(1+x^2)^2$. This function has a maximum of $3\sqrt{3}/16$ at $x=1/\sqrt{3}$, which is also the superior of $x/(1+x^2)^2$ for $x \geq 0$. $-\eta F_{LAD\,\alpha\beta} F_{LAD}{}^{\alpha\beta}$ always has a positive value, which means that R. H. S. in Eq.(27) can't become infinite, strictly. However, if the dynamics is getting run-away (Eq.(26)), the L. H. S. becomes infinite. This term refers to the radiation energy loss due to $dW/dt = m_0\tau_0(dw^\mu/d\tau)(dw_\mu/d\tau) \leq 0$ named the Lamor formula [27]. Since these are inconsistent, $0 \leq -m_0\tau_0(dw^\mu/d\tau)(dw_\mu/d\tau) < \infty$ and $dw^\mu/d\tau$ should be finite. Therefore, the new equation could avoid runaway by infinite light emission. When $\hbar \to 0$, Eq.(27) goes back to the LAD form and the possibility of run-away comes back. Therefore, it is essential to consider radiation reaction with vacuum polarization.

As one check, we will show the calculation result of Eq.(25). The configuration of calculation is that an electron with energy of 200MeV propagates counter to the direction of the laser (see Fig.1). This laser has a Gaussian shape, a linear polarization and an intensity of $1\times 10^{21}\,\text{W/cm}^2$, a wave length of $1\mu m$ and a pulse width of $20\,\text{fsec}$. An electron is $200\times\lambda_{laser}$ from the peak of the laser pulse. In our simulation, the laser propagating direction is $x$, the electric and magnetic fields are in the $y$ and $z$ directions, respectively. Therefore, the shape of electric field is

$$\mathbf{E} = \hat{\mathbf{y}} E_0 \exp\left[-\frac{(\omega_{laser}t - k_{laser}x)^2}{(\omega_{laser}\Delta t)^2}\right] \times \sin(\omega_{laser}t - k_{laser}x), \tag{28}$$

$\hat{\mathbf{y}}$ is the unit vector in the $y$ direction. Since the external laser field has linear polarization,

$$\mathbf{B} = \hat{\mathbf{z}} \frac{E_0}{c} \exp\left[-\frac{(\omega_{laser}t - k_{laser}x)^2}{(\omega_{laser}\Delta t)^2}\right] \times \sin(\omega_{laser}t - k_{laser}x). \tag{29}$$

The result is shown in Fig.2. This is the time evolution of the electron's energy. The red dotted line is calculated by the Landau-Lifshitz (LL) equation, which is one approximation of the LAD equation. The LL equation is as follows [14]:

$$m_0 \frac{d}{d\tau} w^\mu = F_{ex}{}^\mu - eF_{LL}{}^{\mu\nu} w_\nu \tag{30}$$

$$F_{LL}{}^{\mu\nu} = -\frac{\tau_0}{ec^2}\left(\frac{dF_{ex}{}^\mu}{d\tau} w^\nu - \frac{dF_{ex}{}^\nu}{d\tau} w^\mu\right) \tag{31}$$

This equation can be calculated smoothly, because there isn't the second derivative of $w^\mu$ (we call this the Schott term in the LAD equation). Therefore, this is a convenient equation for simulations [20-21, 28-29]. And the blue solid line in Fig.2 is the solution of Eq.(25). Both equations have the same tendency.

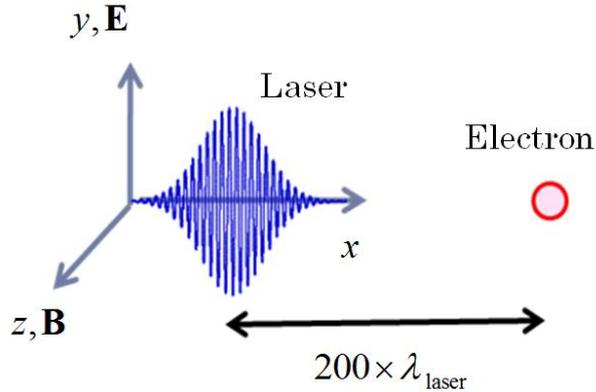

Fig.1 Setting of counter propagating configuration. This laser has an intensity of $1 \times 10^{21}$ W/cm$^2$, a wavelength of $1\mu m$ and a pulse width is $20 fsec$. Initially, the electron is $200 \times \lambda_{laser}$ from the peak of the laser pulse and has an energy of 200MeV ($\gamma = 400$).

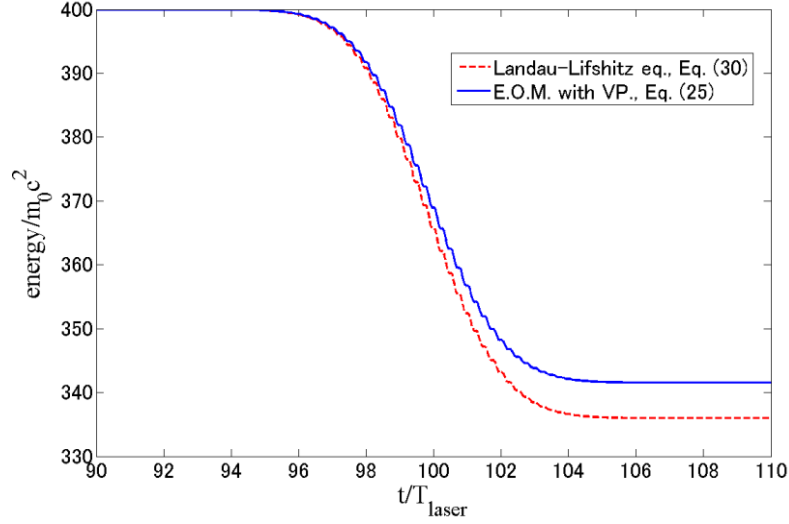

Fig.2  Time evolution of the electron's energy. The electron has an energy of 200MeV ($\gamma = 400$), and emits energy via the process of bremsstrahlung. The blue solid line is the solution of Eq.(25), and the red dotted line is the Landau-Lifshitz equation Eq.(30).

From $t = 95 \times T_{laser}$, the electron emits its kinetic energy. Now, the process in which the electron loses energy is only radiation. This energy loss is the reason why we need to consider interactions under ultraintense lasers. When the electron doesn't emit any electromagnetic wave, the electron maintains its inertial energy $\gamma = 400$. However, the final states are different. The relativistic factor (or normalized electron's energy) $\gamma$ becomes, $\gamma_{Eq.(25)} = 341.5$ and $\gamma_{Eq.(30)} = 336.0$. This difference comes from the part of the Schott term and $1/[1 - 4\alpha^2 \hbar^3 \varepsilon_0 / 45 m_0^4 c^3 (F_{LAD\,\alpha\beta} F_{LAD}^{\alpha\beta})]$, the Landau-Lifshitz method of Eq.(30) and (31) gives a larger estimation for energy loss than the equation of motion of Eq.(25). For now, we cannot determine which equation is correct, this problem has to be solved by experiments.

From Eq.(27), our new equation of motion can always avoid the run-away solution. By using this fact, we are led to the estimation of the radiation reaction in the non-relativistic regime. From our experience, the electron hardly emits light in non-relativistic motion. Therefore, the radiation reaction force is smaller than the external force, we can consider perturbations of Eq.(25) like the method from the LAD equation to the LL equation.

$$F_{LAD}{}^{\mu\nu} \to F_{LL}{}^{\mu\nu} \tag{32}$$

Taking the external field to be relatively weak in this regime (strong external fields lead an electron in the relativistic regime), Eq.(25) becomes the LL equation, because

$$\frac{1}{1-\eta\left(F_{\text{LAD}\,\alpha\beta}F_{\text{LAD}}{}^{\alpha\beta}\right)} \to 1. \tag{33}$$

The non-relativistic limit of Eq.(25) has good convergence to the LL equation.

However, this Heisenberg-Euler Lagrangian is the Lagrangian of a weak field. When applying the equation of motion Eq.(25), it is important how much the LAD field is smaller than the field of the Schwinger limit. The electric field at the Schwinger limit is defined as follows:

$$E_{\text{Schwinger}} = \frac{m_0^2 c^3}{e\hbar} \tag{34}$$

With this notation, the radiation reaction field Eq.(24) becomes,

$$F^{\mu\nu} = \frac{1}{1 - \frac{4\alpha^2 \hbar \varepsilon_0 c}{45 e^2} \frac{c^2 F_{\text{LAD}\,\alpha\beta} F_{\text{LAD}}{}^{\alpha\beta}}{E_{\text{Schwinger}}^2}} F_{\text{LAD}}^{\mu\nu}. \tag{35}$$

Of course, the two factors $4\alpha^2\hbar\varepsilon_0 c/45 e^4 = 5.2\times 10^{-5}$ and $-c^2 F_{\text{LAD}\,\alpha\beta} F_{\text{LAD}}{}^{\alpha\beta}/E_{\text{Schwinger}}^2$ are dimension-less. Stable calculations require the relation of Eq.(33) and the constancy of the LAD field related to the Schwinger limit.

$$E_{\text{Schwinger}}^2 \gg -\frac{4\alpha^2 \hbar \varepsilon_0 c}{45 e^2}\left(c^2 F_{\text{LAD}\,\alpha\beta} F_{\text{LAD}}{}^{\alpha\beta}\right) \approx 10^{-4} \times \left(\mathbf{E}_{\text{LAD}}^2 - c^2 \mathbf{B}_{\text{LAD}}^2\right)\Big|_{\text{at electron}}. \tag{36}$$

Therefore, stable calculations (for example, our calculations shown in Fig. 1 and 2) satisfy Eq.(36), and whether this equation is satisfied or not, will be shown after simulations. When Eq.(36) isn't satisfied, run-away effects will occur. Moreover, the corrections to the Heisenberg-Euler action integral is suppressed by a factor of $\hbar k_\mu \hbar k^\mu / (m_0 c)^2 = 2 k_{\text{laser}\,\mu} k_{\text{radiation}}{}^\mu / (m_0 c)^2$

$$\int d^4 x\, L_{\text{Heisenberg-Euler}}^{\text{1st order}}\left(F_{\alpha\beta}F^{\alpha\beta}, F_{\alpha\beta}{}^* F^{\alpha\beta}\right)$$
$$= -\int d^4 x\, \frac{1}{4\mu_0} F_{\alpha\beta}F^{\alpha\beta} + \int d^4 x\, \frac{\alpha^2 \hbar^3 \varepsilon_0^2}{360 m_0^4 c}\left[4\left(F_{\alpha\beta}F^{\alpha\beta}\right)^2 + 7\left(F_{\alpha\beta}{}^* F^{\alpha\beta}\right)^2\right] + O\left(\left(\frac{\hbar k_{\text{laser}\,\mu} \hbar k_{\text{radiation}}{}^\mu}{m_0^2 c^2}\right)^3\right) \tag{37}$$

[30]. In our calculation, the laser has the wavelength of $1\mu\text{m} = 1.24\text{eV}$. The most effective way in producing radiation reaction is in head-on collisions [Fig.1]. In this case, the photon energy limit of the Radiation is, $\hbar\omega_{\text{radiation}} \sim 10\text{GeV}$. This is the limit of this model.

## 5. Conclusion

In summary, for avoidance of the run-away in the LAD theory, we considered a new model of the radiation reaction with the vacuum polarization and derived a new equation of motion, Eq.(25). The vacuum polarization came from the Heisenberg-Euler Lagrangian [22-24]. The radiation reaction in this paper is based on Dirac's in which, the LAD field is defined as the superposition of the retarded and advanced fields [see Eq.(6)]. This LAD field is a non-source field, that is, it satisfies the same Maxwell's equations obtained from the Heisenberg-Euler Lagrangian. Therefore, considering that these two equations describe the same process, we get the relation in Eq.(15), and the equation of motion,

$$\frac{d}{d\tau}w^\mu = -\frac{e}{m_0\left[1-\eta\left(F_{LAD\,\alpha\beta}F_{LAD}^{\;\;\alpha\beta}\right)\right]}\left(F_{ex}^{\mu\nu}+F_{LAD}^{\;\;\mu\nu}\right)w_\nu. \qquad (38)$$

This new equation of motion is stable, able to describe the energy loss by radiation and has good convergence to the LL equation from the non-relativistic regime. The photon energy limit of the radiation in this model is $\hbar\omega_{radiation} \sim 10\text{GeV}$. The factor

$$\frac{Q}{M} = \frac{e}{m_0\left[1-\eta\left(F_{LAD\,\alpha\beta}F_{LAD}^{\;\;\alpha\beta}\right)\right]} = \frac{e}{m_0} + \frac{\delta e}{m_0} \qquad (39)$$

$$\delta e = e\frac{\eta\left(F_{LAD\,\alpha\beta}F_{LAD}^{\;\;\alpha\beta}\right)}{\left[1-\eta\left(F_{LAD\,\alpha\beta}F_{LAD}^{\;\;\alpha\beta}\right)\right]} \leq 0 \qquad (40)$$

is regarded as the charge to mass ratio[30, 31] with radiation. $\delta e$ is the effect of vacuum polarization, which is named the dress of charge. By rough approximation, this dress depends on the square of the radiation loss $dW/dt = m_0\tau_0(dw^\mu/d\tau)(dw_\mu/d\tau) \leq 0$ [27],

$$\delta e \sim -e \times \frac{2\eta}{e^2c^4}\left(\frac{dW}{dt}\right)^2 = -e\times\left[2.6\times 10^{-20}\times\left(\frac{dW}{dt}\right)^2\right] \qquad (41)$$

Of course, this correction comes from the dress field $M^{\mu\nu}/\varepsilon_0 c$. We should be able to estimate it via the observation of radiation from the electron and analyzing the value of $F_{LAD}$ at the point of an electron. As a next step, we need to proceed to the precision scheme of radiation reaction with a general quantum vacuum model beyond the Heisenberg-Euler Lagrangian.

**Acknowledgement**

This work is partly supported under the auspices of the Japanese Ministry of Education, Culture, Sports, Science and Technology (MEXT) project on "Promotion of relativistic nuclear physics with ultra-intense laser."